\documentclass[
 reprint,
 aps,prl,
 groupedaddress,
 notitlepage,
 superscriptaddress
]{revtex4-1}

\usepackage{color}
\usepackage{float}
\usepackage{graphicx}
\usepackage{amsmath,amstext,amssymb,amsthm}
\usepackage{siunitx}
\DeclareSIUnit{\nothing}{\relax}
\usepackage[hidelinks]{hyperref}
\usepackage[normalem]{ulem}
\usepackage{upgreek}
\usepackage{xspace}

\definecolor{orange}{rgb}{0.59, 0.29, 0.1}
\definecolor{yellow}{rgb}{1.0, 0.7, 0.0}

\newcommand{\overbar}[1]{\mkern 1.5mu\overline{\mkern-1.5mu#1\mkern-1.5mu}\mkern 1.5mu}
\newcommand{\Kbar}{\ensuremath{\overbar{\mathrm{K}}}\xspace}
\newcommand{\Kbarprime}{\ensuremath{\overbar{\mathrm{K}}}$^\prime$\xspace}

\begin{document}

\title{Larger than \texorpdfstring{$\mathbf{80}\,\boldsymbol{\%}$}{80\%} Valley Polarization of Free Carriers\\ in Singly-Oriented Single Layer $\text{WS}_\mathbf{2}$ on $\text{Au(111)}$}

%%%%%%%%%%%%%%%%%%%%%%%%%%%%%%%%%%%%%%%%%%%%%%
\author{H.~Beyer}
\email{hbeyer@physik.uni-kiel.de}
\author{G.~Rohde}
\affiliation{
 Institut f\"ur Experimentelle und Angewandte Physik, Christian-Albrechts-Universit\"at zu Kiel, 24098 Kiel, Germany
}
\author{A.~Grubi\v{s}i\'{c} \v{C}abo}
\affiliation{Department of Physics and Astronomy, Interdisciplinary Nanoscience Center, Aarhus University,
8000 Aarhus C, Denmark}
\author{A.~Stange}
\author{T.~Jacobsen}
\affiliation{
 Institut f\"ur Experimentelle und Angewandte Physik, Christian-Albrechts-Universit\"at zu Kiel, 24098 Kiel, Germany
}
\author{L.~Bignardi}
\author{D.~Lizzit}
\author{P.~Lacovig}
\affiliation{Elettra - Sincrotrone Trieste S.C.p.A., 34149 Trieste, Italy}
\author{C.~E.~Sanders}
\affiliation{
Central Laser Facility, STFC Rutherford Appleton Laboratory, Harwell OX11 0QX, United Kingdom
}
\author{S.~Lizzit}
\affiliation{Elettra - Sincrotrone Trieste S.C.p.A., 34149 Trieste, Italy}
\author{K.~Rossnagel}
\affiliation{
 Institut f\"ur Experimentelle und Angewandte Physik, Christian-Albrechts-Universit\"at zu Kiel, 24098 Kiel, Germany
}
\affiliation{
 Ruprecht-Haensel-Labor, Christian-Albrechts-Universit\"at zu Kiel und Deutsches Elektronen-Synchrotron DESY,
 24098 Kiel und 22607 Hamburg, Germany
}
\affiliation{
 Deutsches Elektronen-Synchrotron DESY, 22607 Hamburg, Germany
}
\author{P.~Hofmann}
\affiliation{Department of Physics and Astronomy, Interdisciplinary Nanoscience Center, Aarhus University,
8000 Aarhus C, Denmark}

\author{M.~Bauer}
\homepage{http://www.physik.uni-kiel.de/en/institutes/bauer-group}
\affiliation{
 Institut f\"ur Experimentelle und Angewandte Physik, Christian-Albrechts-Universit\"at zu Kiel, 24098 Kiel, Germany
}

%%%%%%%%%%%%%%%%%%%%%%%%%%%%%%%%%%%%%%%%%%%%%%

\date{\today}

\begin{abstract}
We employ time- and angle-resolved photoemission spectroscopy to study the spin- and valley-selective photoexcitation and dynamics of free carriers at the \Kbar and \Kbarprime points in singly-oriented single layer $\text{WS}_2/\text{Au}(111)$. Our results reveal that in the valence band maximum an ultimate valley polarization of free holes of \SI{84}{\percent} can be achieved upon excitation with circularly polarized light at room temperature. Notably, we observe a significantly smaller valley polarization for the photoexcited free electrons in the conduction band minimum. Clear differences in the carrier dynamics between electrons and holes imply intervalley scattering processes into dark states being responsible for the efficient depolarization of the excited electron population.
\end{abstract}

\maketitle

\begin{figure}
        \includegraphics[width=1\linewidth]{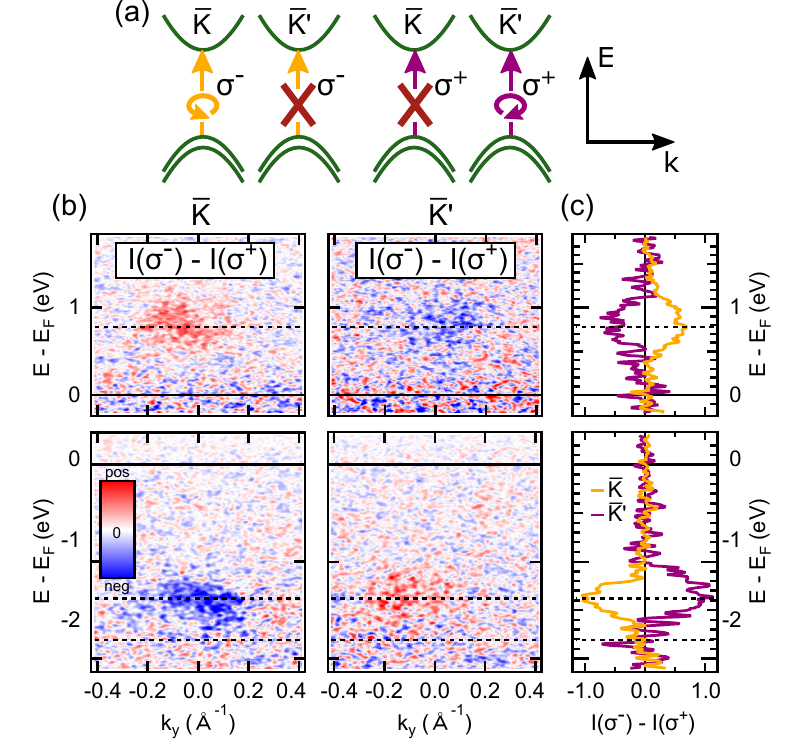}
        \caption{Detection of valley-selective excitation of the WS$_2$ layer using circularly polarized light. (a) Schematic illustration of the optical selection rules of single layer WS$_2$ for valley-selective excitation at \Kbar and \Kbarprime. (b) Difference photoemission intensity maps at \Kbar and \Kbarprime obtained from trARPES spectra recorded upon excitation with $\sigma^-$ and $\sigma^+$ polarized pump pulses at $\Delta t =\SI{50}{\femto\s}$. The top (bottom) panels show conduction (valence) band data. (c) Comparison of normalized difference EDCs at \Kbar and \Kbarprime derived from the data shown in~(b). The signal was integrated over a momentum window of \SI{0.35}{\per \angstrom} and normalized to the VBM peak value.
        }
	    \label{fig:DynamikVergleich}
    \end{figure}

\noindent Semiconducting single layer transition metal dichalcogenides (SL~TMDCs) are promising platforms for future opto-valleytronic and opto-spintronic applications \cite{Wang2012, Butler2013, Xu2014, Mak2016}. The remarkable properties of these materials arise from the presence of a direct band gap at the \Kbar and \Kbarprime valleys in combination with a lack of structural inversion symmetry and strong spin-orbit coupling. Valley and spin degrees of freedom are strongly coupled so that a valley-selective excitation of spin-polarized carriers upon absorption of circularly polarized light becomes possible \cite{Xiao2012, Cao2012, Zeng2012, Mak2012} [Fig.~\ref{fig:DynamikVergleich}(a)].\\
Experimentally, the unique properties of SL~TMDCs were studied predominantly by all-optical techniques providing particular insights into the intriguing exciton physics of these materials \cite{Mak2010, Splendiani2010, Chernikov2015, Yang2015, Rivera, Christiansen2017}. For the investigation of free carrier processes it is advantageous to alternatively apply photoemission techniques, which can provide the energy, momentum and spin sensitivity required to map out the ground and excited state electronic band structure and their properties \cite{Riley2014, Hein2016, Bruix2016, Wallauer2016, Waldecker2017}.
The direct study of SL~TMDCs by photoemission spectroscopy relies, however, on high quality TMDC layers with typical sizes in the mm$^2$-regime. Bottom-up growth techniques allow for the production of such types of samples and were, for instance, successfully applied for the preparation of SL~TMDCs on single crystalline noble metal substrates \cite{Grønborg2015, Dendzik2015}. Previous angle-resolved photoemission spectroscopy (ARPES) and time-resolved ARPES (trARPES) studies of such samples revealed insights into the electronic structure and the ultrafast free carrier dynamics \cite{Grønborg2015, GrubisicCabo2015}. Furthermore, it was possible to demonstrate optical control of the spin and valley degrees of freedom using circularly polarized light \cite{Ulstrup2016}. A critical drawback of these samples is, however, the presence of mirror domains \cite{Pulkin2015}, which show an inversion of the \Kbar and \Kbarprime points. As photoemission experiments intrinsically average over these domains, effects due to an optically induced spin and valley selectivity become reduced or completely masked. A quantitative interpretation of such data is therefore difficult or even impossible.\\
This Letter reports on a trARPES study of a singly-oriented layer of WS$_2$ epitaxially grown on $\text{Au}(111)$. The single orientation character of the sample with a maximum of \SI{5}{\percent} contribution of mirror domains was demonstrated in a comprehensive study on the structural properties \cite{Bignardi2019}. 
This unique property makes it possible to gain quantitative information on the valley selectivity of free carrier excitation using circularly polarized light.
We show that in the valence band maximum (VBM) a valley polarization of free holes of \SI{0.84(16)}{} can be generated. Remarkably, the free electron valley polarization in the conduction band minimum (CBM) is lower with a value of \SI{0.56(16)}{}. We consider intervalley scattering processes between \Kbar and \Kbarprime being responsible for this reduction, which are strongly enhanced in the conduction band (CB) due to an almost vanishing spin splitting. Differences in the observed depopulation rates of excited carriers between CBM and VBM support this interpretation.\\
The singly-oriented SL WS$_2$/$\text{Au}(111)$ sample was grown with a coverage of about \SI{45}{\%} at the SuperESCA beamline of the Elettra Synchrotron radiation facility in Trieste \cite{Bignardi2019}. The sample was transported to the Kiel trARPES system in an evacuated tube and cleaned by laser annealing under ultrahigh vacuum (UHV) conditions using \SI{400}{nm} laser pulses ($\lesssim$\,\SI{50}{fs} pulse width) at an incident fluence of several mJcm$^{-2}$. The procedure was applied until ARPES spectra of the characteristic band structure of WS$_2$ did not show any further changes. TrARPES experiments were performed using the output of a \SI{7.2}{\kilo\Hz} Ti:sapphire multipass amplifier. Near-resonant excitation at the \Kbar and \Kbarprime points of WS$_2$ was achieved at an incident fluence of $F \approx \SI{300}{\micro\J\per\square\cm}$ using \SI{2.10}{\eV} (\SI{590}{\nano\m}) pump pulses generated in a non-collinear optical parametric amplifier. The polarization of the pump pulses was adjusted with a zero-order quarter-wave plate (QWP). A Stokes polarimeter was used for the quantitative characterization of the polarization state of the pulses. ARPES probe spectra were recorded with an energy resolution of \SI{390}{\milli\eV} using p-polarized \SI{22.1}{\eV} pulses delivered from a high harmonic generation (HHG) source \cite{Eich2014} and using a hemispherical analyzer. Pump and probe pulses were focused almost collinearly at near-normal incidence onto the sample. Cross correlation measurements at the sample position yielded a time resolution of \SI{130}{\femto\s} (FWHM). All experiments were performed at a pressure of \SI{3 e-10}{\milli \bar} and a sample temperature of \SI{300}{\K}.

\begin{figure}
        \includegraphics[width=1\linewidth]{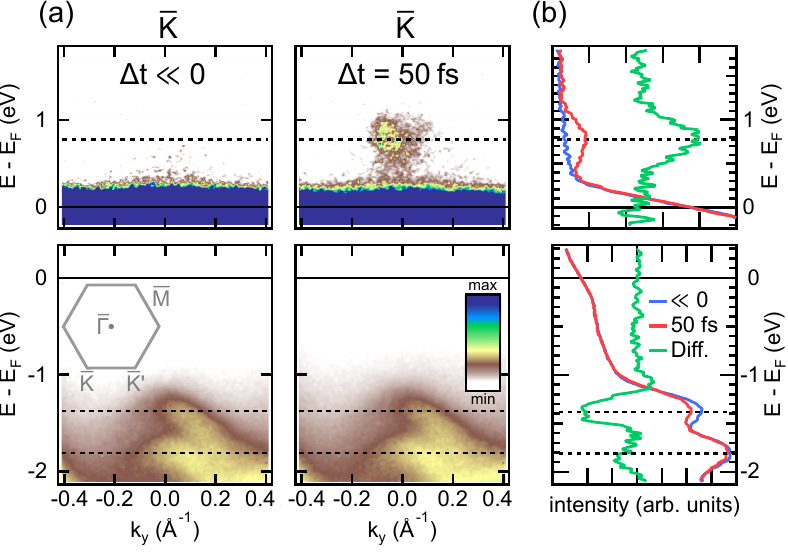}
        \caption{Time-resolved ARPES data of $\text{WS}_2/\text{Au}(111)$ taken at \Kbar for excitation with linearly polarized \SI{2.10}{eV} pump pulses. (a) ARPES snapshots recorded before optical excitation (left) and at $\Delta t=$ \SI{50}{\femto\s} (right). Top (bottom) panels show the conduction (valence) band signal. Color scales have been adjusted separately for top and bottom panels to account for the significantly different photoemission intensities from CB and VB. The inset in the bottom left panel shows the hexagonal Brillouin zone of SL $\text{WS}_2$. (b) Comparison of energy distribution curves (EDCs) derived from the ARPES data shown in (a). Photoemission intensity was integrated over a momentum window of \SI{0.2}{\per \angstrom}. The green line shows the difference calculated from the two EDCs.}
    	\label{fig:StatischeARPES}
\end{figure}

Figure \ref{fig:StatischeARPES}(a) compares trARPES data of the sample around \Kbar before (negative pump-probe delay $\Delta t$) and during ($\Delta t=\text{\SI{50}{\femto \s}}$) the optical excitation with linearly polarized \SI{2.10}{\eV} pump pulses. In both spectra one can clearly distinguish the spin-split upper and lower $\text{WS}_2$ valence bands (UVB/LVB) below $E_{\mathrm {F}}$ (bottom panels).
The additional signal at $\Delta t=\text{\SI{50}{\femto \s}}$ for energies $E>E_{\mathrm {F}}$ (top panels) results from the transient population of the conduction band at the CBM due to the optical excitation.
Energy distribution curves (EDCs) derived from the data in Fig.\ \ref{fig:StatischeARPES}(a) are shown in Fig.\ \ref{fig:StatischeARPES}(b). A difference spectrum calculated from the EDCs (green line) furthermore uncovers a transient depletion of the carrier population near the UVB maximum. A finite, but much weaker depletion is also visible for the LVB.

Quantitative analysis of the spectra yields a direct gap of $E_{\mathrm{gap}}=\SI{2.06(7)}{\eV}$ and an energy splitting between UVB and LVB of $\Delta E_{\text{VB}}=\SI{440(70)}{\milli\eV}$ \cite{Supplemental}\nocite{Berry1977, Schaefer2017}. Both values are in very good agreement with earlier experiments \cite{Dendzik2015, Ulstrup2016, Eickholt2018} and indicate a resonant excitation between upper VBM and CBM at the used photon energy of \SI{2.10}{\eV}. The observed depletion of the LVB can be associated with a transition into gap states near $E_{\mathrm {F}}$ \cite{Gong2014, Liu2013}.

A recent trARPES study of the semiconducting bulk TMDC 2\textit{H}-MoSe$_2$ reported additionally on the observation of transient excitonic signatures \cite{Buss17}. In our case, screening due to the free carriers of the supporting gold substrate efficiently suppresses the formation of bound excitons \cite{Ugeda2014}. Furthermore, the presented experiments are performed at an excitation density well above the threshold for an excitonic Mott transition in SL WS$_2$ \cite{Chernikov2015}.

\begin{figure}
        \includegraphics[width=1\linewidth]{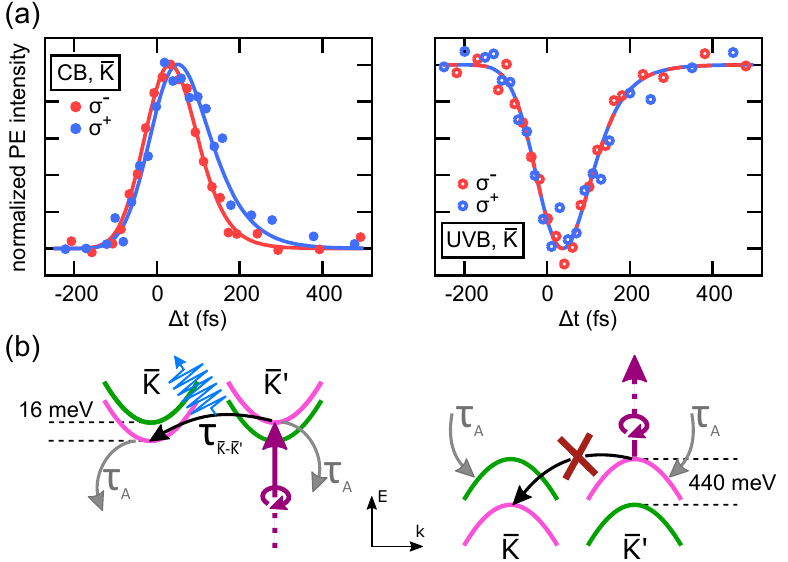}
        \caption{Intervalley scattering of free carriers in valence and conduction band. (a) Comparison of the temporal evolution of normalized PE intensities in the CB (left) and the UVB (right) at \Kbar for excitation with (predominantly)  $\sigma^-$ and $\sigma^+$ circularly polarized light. The solid lines are the results of a fit of a rate equation model to the experimental data as described in the text. For better comparison, the data is normalized to the maximum/minimum transient intensity, respectively. (b)~Schematic illustration of intervalley scattering from \Kbarprime to \Kbar for direct photoexcitation at \Kbarprime (excitation with $\sigma^+$ circularly polarized light). Phonon emission (blue arrow) accounts for energy conservation in the intervalley scattering process in the CB. The time constants $\uptau_{A}$ and $\uptau_{\text{\Kbar}-\text{\Kbarprime}}$ describe the decay of excited carriers according to the used rate equation model \cite{Supplemental}.} 
    	\label{fig:DichroismusDynamik}
    \end{figure}

Photoinduced valley-selectivity within the WS$_2$ layer is demonstrated by comparing transient ARPES spectra at \Kbar (\Kbarprime) recorded \SI{50}{fs} after excitation with right ($\sigma^+$) and left ($\sigma^-$) circularly polarized pump pulses, respectively.
The specific delay was chosen so that the transient intensity becomes maximum, see Fig.\ \ref{fig:DichroismusDynamik}(a).
Difference intensity maps generated from these spectra are shown in Fig.~\ref{fig:DynamikVergleich}(b). The data confirms the presence of a strong circular dichroism both in the CB and in the valence band (VB). The contrast is inverted between the \Kbar and \Kbarprime points, as expected from the optical selection rules. In the VB a dichroism is only observed in the UVB, but is absent in the LVB, as can be seen particularly clearly in the difference EDCs shown in Fig.~\ref{fig:DynamikVergleich}(c).

The experimental data presented so far confirm the qualitative findings of a related study on a SL WS$_{2}$/Ag(111) sample that exhibited a preferential, but not single domain orientation \cite{Ulstrup2016}. In the following, we will show that the single orientation character of our sample allows also for a quantitative determination of the valley polarization that ultimately can be generated upon optical excitation.

    \begin{figure}
        \includegraphics[width=1\linewidth]{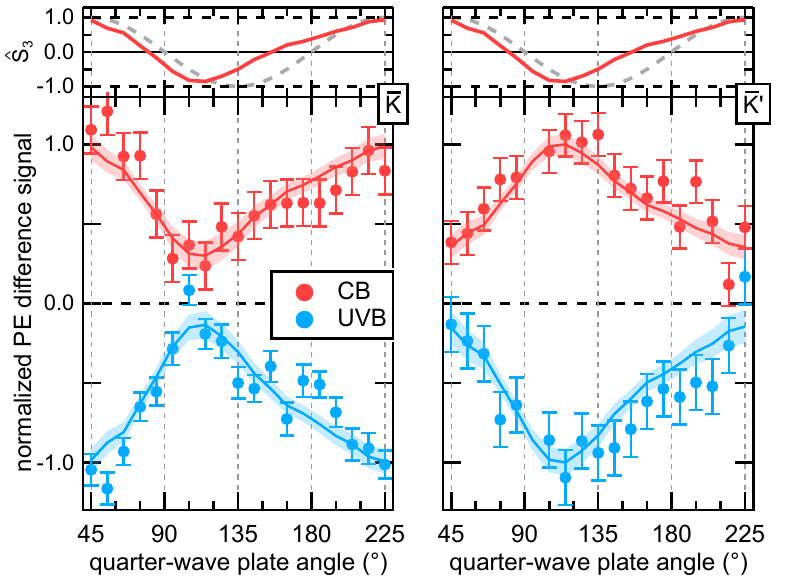}
        \caption{Photoemission (PE) signal of CB and UVB at \Kbar and \Kbarprime as a function of the angle of the quarter-wave plate in the pump beam ($\Delta t=\SI{50}{\femto\s}$). For the evaluation of the UVB data, an equilibrium state spectrum ($\Delta t \ll 0$) was subtracted from the excited state spectrum. CB (UVB) traces are normalized to maximum (minimum) PE signal. The error bars of the experimental data account for the uncertainties in determining the signal background not originating from the valley population. The solid lines are the results of the fits of Eq.\ \ref{eq:1} to the experimental data. The errors in the fits (shaded areas) account for error propagation of the fitting results and the uncertainties in determining $\hat{S}_{3}$. The top panel displays the normalized Stokes parameter $\hat{S}_{3}$ determined from the Stokes polarimeter measurements of the pump pulse \cite{Supplemental} (red line) in comparison to an ideally polarized pump pulse (dashed gray line).}
	    \label{fig:DichroismusStatisch}
    \end{figure} 
    
In the further investigations we performed pump polarization scans with the angle of the QWP in the pump beam varied over a range of \ang{180} in steps of \ang{10}. Results are summarized in Fig.\ \ref{fig:DichroismusStatisch}, which shows normalized integral photoemission intensities for $\Delta t = \SI{50}{fs}$ of the CBM (red) and the upper VBM (blue) as a function of the QWP angle. As expected for a dichroic response, we observe distinct maxima and minima as the circular polarization state is changed. The inversion of the traces at \Kbar and \Kbarprime is in agreement with the valley selectivity of the excitation process shown above. Notably, the traces exhibit a clear asymmetry with respect to the QWP angle, shifting the extrema expected at \SI{135}{\degree} by approximately $\SI{-20}{\degree}$.  
The polarization scan allows to quantify the circular dichroism $D= (I_{\text{max}} - I_{\text{min}})/ (I_{\text{max}} + I_{\text{min}})$ in the photoemission signal with $I_{\text{max}}$ and $I_{\text{min}}$ being the maxima and minima in the photoemission signal, respectively. The analysis yields circular dichroism values of $D=0.7$ for the UVB and $D=0.5$ for the CB. 
Surprisingly, the circular dichroisms in the photoemission signal from UVB and CB clearly differ.

Further quantitative analysis of the data relies on a detailed characterization of the changes in the circular polarization state of the pump pulse as the QWP angle is changed. Measurements were performed with the Stokes polarimeter and are presented and discussed in detail in the Supplemental Material \cite{Supplemental}. The upper panels of Fig.~\ref{fig:DichroismusStatisch} show the evaluated normalized Stokes parameter $\hat{S}_{3}$ of the pump pulses at the sample position as a function of the QWP angle. We observe a distinct asymmetry in the data, which can be traced back to the reflection from the final deflection mirror mounted inside the UHV chamber \cite{Supplemental}. Additionally, the quantitative analysis of the data yields a maximum absolute value for the normalized Stokes parameter $\hat{S}_{3}$ of \SI{0.9}{}, i.e.\xspace, it is not possible in this configuration to observe a circular dichroism of \SI{100}{\percent}.  
A comparison with the ARPES data in Fig.\ \ref{fig:DichroismusStatisch} implies that part of the observed peculiarities in the photoemission polarization scans directly reflect the circular polarization state of the pump pulse.

The Stokes polarimeter results enable us to evaluate the fraction of carriers $p$ excited according to the optical selection rules and from this the degree of valley polarization $P=(2\cdot p-1)$.
For a given fraction $f$ of preferentially oriented domains, the changes in the integrated photoemission signal $I_{\Kbar}$ at the \Kbar point during a pump polarization scan can be described by the relation \cite{Supplemental}
\begin{align}
     I_{\Kbar} \propto \ &c \cdot f \cdot p \,+\, (1-c) \cdot (1-f) \cdot p \notag \\ 
     +& \, c \cdot (1-f) \cdot (1-p)  \,+\, (1-c) \cdot f \cdot (1-p) .  
\label{eq:1}
\end{align}

Here, $c = 0.5 \cdot \hat{S}_{3} + 0.5$ denotes the degree of circular polarization of the pump pulse with $c = +1$ ($c = 0$) corresponding to purely $\sigma^-$ ($\sigma^+$) polarized light.

The solid lines in Fig.\ \ref{fig:DichroismusStatisch} show fits of Eq.\ \ref{eq:1} to the experimental data with $p$ being the only free fitting parameter yielding $p=\SI{0.92(11)}{}$ for the UVB and $p=\SI{0.78(11)}{}$ for the CB. The value for $c$~was determined from the Stokes polarimeter data and $f$ was set here to $f=1$ accounting for a perfectly oriented WS$_2$ layer.
We conclude that upon excitation with purely circularly polarized laser pulses an almost perfect valley selective hole population in the UVB can be prepared.
The observed value of $p$ for the UVB results in a valley polarization $P=\SI{0.84(16)}{}$. Note that in the presence of mirror domains this value can only increase. For the limiting case of a \SI{5}{\percent} contribution of mirror domains ($f = 0.95$) \cite{Bignardi2019} we obtain $P=\SI{0.94}{}$ ($p=\SI{0.97}{}$). 
For comparison, for the valley polarization of \textit{A} excitons in semiconducting SL TMDCs, theory predicts $P=0.90$, which in this case is limited by coherent intervalley coupling \cite{Berghauser2018}.

The analysis of the CB data yields $P=\SI{0.56(16)}{}$ ($P=\SI{0.62}{}$  for the limiting case of $f=0.95$). In agreement with the observed differences in the circular dichroism, these values are significantly smaller than what we evaluated for the UVB. The analysis of time-resolved photoemission data presented in the following section provides further insights into the origin of this difference.

Figure \ref{fig:DichroismusDynamik}(a) compares normalized photoemission intensity transients for CBM and upper VBM deduced from trARPES data at the \Kbar point with the QWP set to \SI{45}{\degree} and \SI{135}{\degree}, i.e., for excitation with predominantly $\sigma^-$ and $\sigma^+$ polarized pump pulses, respectively. Note that despite the optical selection rules we observe for both cases a finite transient signal at \Kbar due to the not perfectly circularly polarized light of the pump pulse, potential contributions from mirror domains, and a value of $p < 1$.
The overall temporal evolution of the transients reflects the excited carrier population and relaxation dynamics, with the latter one being largely governed by Auger-type processes due to interaction with charge carriers in the gold substrate \cite{GrubisicCabo2015}. Notably, for the CBM we observe clear differences in the temporal evolution for $\sigma^+$ and $\sigma^-$ excitation. This implies a distinct delay in the population of the CB valley at \Kbar for the case of a predominant photoexcitation at \Kbarprime (using $\sigma^+$ light). We conclude that for this excitation scenario the CB valley at \Kbar becomes in large part populated indirectly, and therefore delayed, by intervalley scattering from \Kbarprime. This indirect excitation additionally reduces the overall valley selectivity during the finite duration of the excitation process in our experiment. In contrast, the temporal evolution of the UVB transients remains unchanged upon switching from $\sigma^+$ to $\sigma^-$ excitation. 
In both cases the transient hole population in the UVB directly results from the photoexcitation process. If present at all, contributions from intervalley scattering processes are negligibly slow \cite{Mai2014, Mai2014_1}.

The distinct differences in the spin-orbit splitting between VB and CB, as illustrated in Fig.\ \ref{fig:DichroismusDynamik}(b), can account for the differences in the observed dynamics. The small spin-orbit splitting at the CBM of only \SI{16}{\milli\eV}  \cite{Eickholt2018} opens up spin-conserving intervalley scattering channels for photoexcited electrons from \Kbarprime into the energetically lower dark states at \Kbar via phonon emission processes, as indicated by the black arrow in Fig.\ \ref{fig:DichroismusDynamik}(b) \cite{Jin2014, Hinsche2017}. For photoexcited holes at the upper VBM this channel is efficiently blocked due to the large spin-orbit splitting of \SI{440}{\milli\eV}, which considerably exceeds the maximum phonon energy in the system. Therefore, we propose that this spin-conserving intervalley scattering channel is responsible for the observed accelerated depopulation at the directly photoexcited CBM and the reduction in the valley polarization in the CB in comparison to the VB.

The intervalley scattering rate can be determined from a rate equation analysis of the photoemission intensity transients of the CB and VB \cite{Supplemental}. Fits of the rate equation model to the experimental data are added for comparison as solid lines in Fig.~\ref{fig:DichroismusDynamik}(a). 
The fits to the VB data yield a characteristic depopulation time constant $\uptau_{A}=\SI{60(20)}{\femto\s}$, independent of whether the direct ($\sigma^-$) or the indirect ($\sigma^+$) excitation scenario is considered. This value is in good quantitative agreement with results reported for other SL TMDCs on noble metal substrates \cite{GrubisicCabo2015,Ulstrup2016}
and can be associated with the population decay due to Auger-type interaction processes with carriers in the gold substrate. 
$\uptau_{A}$ is used as an input for the fits to the CB data, making the intervalley scattering time constant, $\uptau_{\text{\Kbar}-\text{\Kbarprime}}$, the only free fitting parameter. These fits give a value of $\uptau_{\text{\Kbar}-\text{\Kbarprime}} = \SI{150(50)}{\femto\s}$. Notably, this value agrees well with the typical timescales on the order of \SI{100}{\femto\s} predicted from theory for the formation of momentum forbidden intervalley dark excitons in W-based SL TMDCs due to electron-phonon interaction \cite{Selig2018, Selig2016}.

In summary, our trARPES study of SL WS$_2$/$\text{Au}(111)$ shows a very high valley polarization in the excited state photoemission signal. This observation confirms the absence of structural mirror domains in the studied WS$_2$ layer, as was shown in a previous study of the investigated sample \cite{Bignardi2019}.
On a closer look, we find that the valley polarization of free holes in the upper VB considerably exceeds the value for the free electrons in the CB. Substantial differences in the transient evolution of the CB intensity point to a coupling channel between \Kbar and \Kbarprime that is not available for the excited carriers in the VB. This behavior can be explained by the different spin-orbit splitting of VB and CB. The herein reported valley polarization of \SI{84}{\%} at room temperature shows that free hole excitations in SL WS$_2$ can be particularly attractive for future opto-spintronic applications.\\

\begin{acknowledgments}
This work was supported by the German Research Foundation (DFG) through project BA 2177/10-1.
We gratefully acknowledge funding from VILLUM FONDEN through the Centre of Excellence for Dirac Materials (Grant.\ No.\ 11744) and the Danish Council for Independent Research, Natural Sciences under the Sapere Aude program (Grant No.\ DFF-4002-00029).
\end{acknowledgments}

\bibliography{bibliography}

\end{document}

\title{Supplemental Material for\texorpdfstring{\\}{ }``Larger than \texorpdfstring{$\mathbf{80}\,\boldsymbol{\%}$}{80\%} Valley Polarization of Free Carriers\\ in Singly-Oriented Single Layer $\text{WS}_\mathbf{2}$ on $\text{Au(111)}$''}

%%%%%%%%%%%%%%%%%%%%%%%%%%%%%%%%%%%%%%%%%%%%%%
\author{H.~Beyer}
\email{hbeyer@physik.uni-kiel.de}
\author{G.~Rohde}
\affiliation{
 Institut f\"ur Experimentelle und Angewandte Physik, Christian-Albrechts-Universit\"at zu Kiel, 24098 Kiel, Germany
}
\author{A.~Grubi\v{s}i\'{c} \v{C}abo}
\affiliation{Department of Physics and Astronomy, Interdisciplinary Nanoscience Center, Aarhus University,
8000 Aarhus C, Denmark}
\author{A.~Stange}
\author{T.~Jacobsen}
\affiliation{
 Institut f\"ur Experimentelle und Angewandte Physik, Christian-Albrechts-Universit\"at zu Kiel, 24098 Kiel, Germany
}
\author{L.~Bignardi}
\author{D.~Lizzit}
\author{P.~Lacovig}
\affiliation{Elettra - Sincrotrone Trieste S.C.p.A., 34149 Trieste, Italy}
\author{C.~E.~Sanders}
\affiliation{
Central Laser Facility, STFC Rutherford Appleton Laboratory, Harwell OX11 0QX, United Kingdom
}
\author{S.~Lizzit}
\affiliation{Elettra - Sincrotrone Trieste S.C.p.A., 34149 Trieste, Italy}
\author{K.~Rossnagel}
\affiliation{
 Institut f\"ur Experimentelle und Angewandte Physik, Christian-Albrechts-Universit\"at zu Kiel, 24098 Kiel, Germany
}
\affiliation{
 Ruprecht-Haensel-Labor, Christian-Albrechts-Universit\"at zu Kiel und Deutsches Elektronen-Synchrotron DESY,
 24098 Kiel und 22607 Hamburg, Germany
}
\affiliation{
 Deutsches Elektronen-Synchrotron DESY, 22607 Hamburg, Germany
}
\author{P.~Hofmann}
\affiliation{Department of Physics and Astronomy, Interdisciplinary Nanoscience Center, Aarhus University,
8000 Aarhus C, Denmark}

\author{M.~Bauer}
\homepage{http://www.physik.uni-kiel.de/en/institutes/bauer-group}
\affiliation{
 Institut f\"ur Experimentelle und Angewandte Physik, Christian-Albrechts-Universit\"at zu Kiel, 24098 Kiel, Germany
}

%%%%%%%%%%%%%%%%%%%%%%%%%%%%%%%%%%%%%%%%%%%%%%

\date{\today}

\maketitle

\section{Band gap and valence band splitting}

\begin{figure}
        \includegraphics[width=0.99\linewidth]{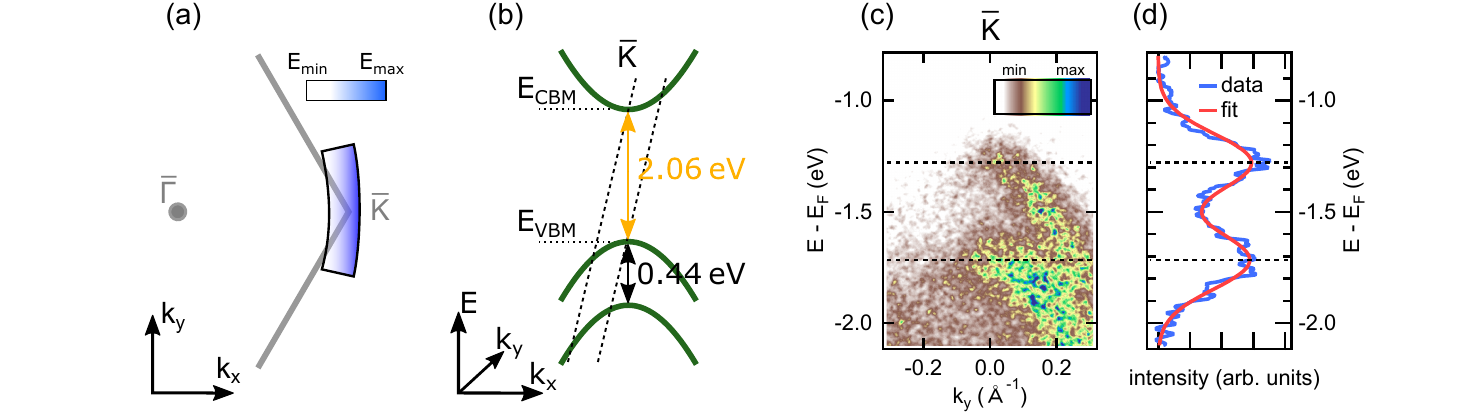}
        \caption{ARPES data of $\text{WS}_2/\text{Au}(111)$ taken at \Kbar for an energy-momentum cut through the VBM. (a) Energy-momentum cut with the analyzer slit oriented perpendicular to the \Gammabar\Kbar direction. The color coded area illustrates the relation between $k_\text{x}$ and $E_{\text{kin}}$ for a cut as investigated in the present study. (b) Schematic illustration of the energy-momentum cuts (dashed lines) through VBM and CBM chosen for the ARPES spectra shown in Fig.\ 1(b) of the main text and in Fig.\ \ref{fig:BandGap}(c) of the supplemental material, respectively. Note that the band structure is displayed along $k_x$, which is perpendicular to the entrance slit of the analyzer. (c) ARPES spectrum for an energy-momentum cut through the VBM showing the spin split upper and lower VB. (d) Background subtracted EDC integrated over a momentum window of \SI{0.1}{\per \angstrom}. The red line shows the result of a fit of two Gaussians to the EDC.}
    	\label{fig:BandGap}
\end{figure}

\noindent For this experiment, the detection geometry is chosen to be perpendicular to the high symmetry direction \Gammabar\Kbar, as depicted in Fig.\ \ref{fig:BandGap}(a). Therefore, in order to evaluate the electronic band gap of the $\text{WS}_2$ layer from the experimental data, one has to account for the tilt of the detection plane in energy-momentum space mapped in the trARPES experiment, as $k_\text{x} \propto \sqrt{E_{\text{kin}}}$ with $k_\text{x}$ being the electron wave vector along \Gammabar\Kbar [see Fig. \ref{fig:BandGap}(a)] and $E_{\text{kin}}$ the kinetic energy of the photoemitted electrons. The energies of valence band maximum (VBM) and conduction band minimum (CBM) must be determined separately from two different energy-momentum cuts as illustrated in Fig.\ \ref{fig:BandGap}(b). 
Figures \ref{fig:BandGap}(c) and (d) show the ARPES spectrum and the corresponding energy distribution curves (EDC) at \Kbar for an energy-momentum cut through the VBM. Figure 1 of the main text shows the data for an energy-momentum cut through the CBM. The energies of VBM and CBM were determined
from the peak maxima of Gaussian fits to the EDCs [see solid red line in Fig.~\ref{fig:BandGap}(d)]. This analysis yields energies of $E_\text{CBM}-E_{\mathrm {F}}=\SI{0.78(5)}{\eV}$ for the CBM and $E_\text{VBM}-E_{\mathrm {F}}=\SI{-1.28(5)}{\eV}$ for the VBM resulting in an energy of the direct band gap of $E_\text{gap}=\SI{2.06(7)}{\eV}$. 
Note that an evaluation of the energies of VBM and CBM from a single energy-momentum cut would result in an overestimation of $E_\text{gap}$ by approx.\ \SI{+100}{\milli\eV}. 
The energy $E_\text{LVB}$ of the maximum of the lower valence band was determined from the data shown in Fig.\ \ref{fig:BandGap}(c). The fit yields a value of $E_\text{LVB}-E_{\mathrm {F}}=\SI{-1.72(5)}{\eV}$ resulting in a valence band splitting of $\Delta E_\text{VB}=\SI{0.44(7)}{\eV}$.

\section{Characterization of the polarization state of the pump pulses}

\noindent For a complete characterization of the polarization state of the pump pulses we measured the Stokes vector $\Vec{S}$ using a Stokes polarimeter \cite{Berry1977, Schaefer2017}. The components $S_{i},\, i \in \lbrace 1,2,3 \rbrace$, of the Stokes vector are referred to as Stokes parameters and quantify linear ($S_1$ and $S_2$) and circular ($S_3$) contributions to the polarization state of the light. Unlike the Jones vector, the Stokes vector is defined by intensity differences, i.e.\xspace, the Stokes parameters can be interpreted as the difference in intensity measured for orthogonal polarization states behind an ideal polarizer, e.g.\xspace, $S_3 = I(\sigma^-) - I(\sigma^+)$. 
With $S_0$ denoting the total intensity of the light, one can introduce the normalized stokes parameters $\hat{S}_{i} = S_i/S_0 ,\, i \in \lbrace 1,2,3 \rbrace$, yielding values of $1$ or $-1$ in the case that the light is fully polarized along the respective orientations. Based on the normalized Stokes parameter $\hat{S}_{3}$ we define the degree of circular polarization $c = (0.5 \cdot \hat{S}_{3} + 0.5)$, which is proportional to the intensity of the $\sigma^-$ polarized light. $c$ is used as an input to evaluate the normalized photoemission signal according to Eq.\ (1) in the main text.

\begin{figure}
        \includegraphics[width=1.0\linewidth]{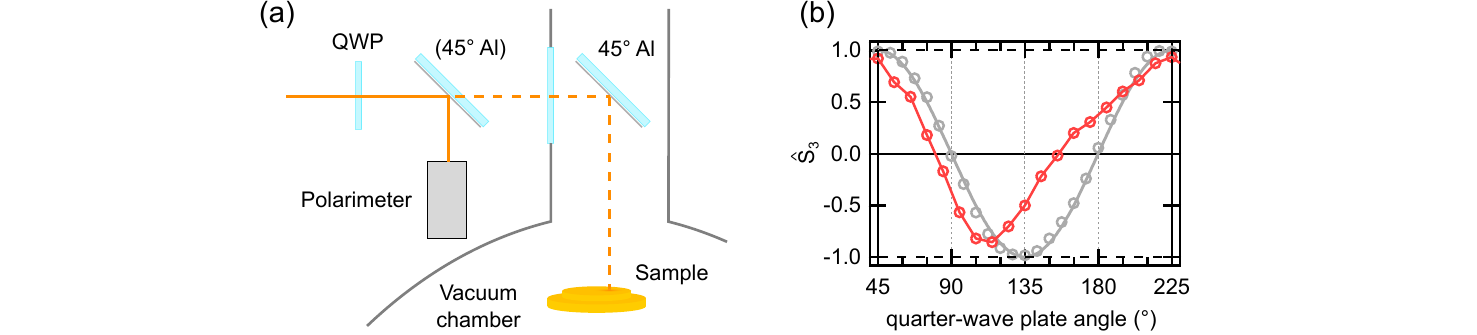}
        \caption{Characterization of the polarization state of the pump pulses. (a) Experimental scheme with replicated beam path for the measurement of the Stokes vector. After passing a quarter-wave plate (QWP), the pump pulses are reflected at an angle of $\SI{45}{\degree}$ from an aluminum mirror and are finally analyzed by the Stokes polarimeter. (b) Normalized Stokes parameter $\hat{S}_{3}$ as a function of the angle of the QWP with (red) and without (gray) the aluminum mirror placed in the beam path.
        }
    	\label{fig:Stokes}
\end{figure}

The Stokes polarimeter could not be mounted inside the UHV chamber at the sample position. Measurements were instead performed with a copy of the pump beam path as schematically illustrated in Fig.\ \ref{fig:Stokes}(a).
We observe the minimum value of $\hat{S}_{3}$ at an angle of the QWP shifted by $\SI{-20}{\degree}$ and the amplitude being reduced to a value of $\SI{0.9}{}$ in comparison to the sinusoidal relation expected for an ideal configuration for the preparation of fully circularly polarized light pulses [see red and gray curves in Fig.\ \ref{fig:Stokes}(b)]. These deviations result from the different reflectances and phase shifts for the s- and p-components of the light pulse upon $\SI{45}{\degree}$-reflection from the final aluminum mirror in the beam path. The design of the experimental setup does not allow for the removal of this aluminum mirror and direct illumination of the sample. Note that the fused silica vacuum window is non-birefringent and therefore does not alter the polarization state.

\section{Derivation of equation (1) in the main text}

\begin{figure}
        \includegraphics[width=1.0\linewidth]{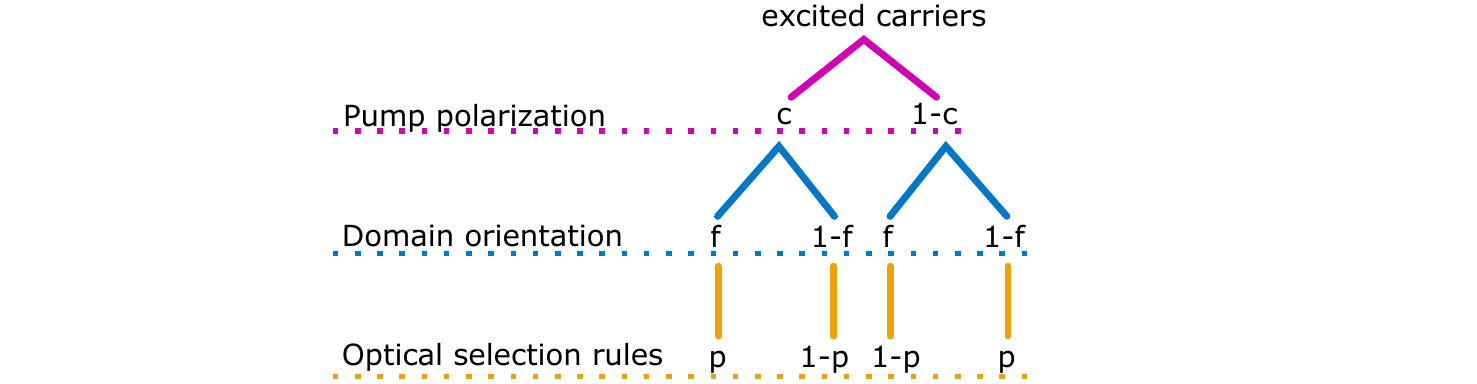}
        \caption{Schematic illustration of the different contributions to the observed excited state intensity as quantitatively described by Eq.\ (1) in the main text. According to the circular polarization $c$ of the pump pulse, carriers get excited at either \Kbar or \Kbarprime. Contributions from domains rotated by \SI{0}{\degree} (\SI{60}{\degree}) are proportional to $f$ ($1-f$) of the total excited state intensity. The fraction of carriers behaving (not) according to the optical selection rules is given by $p$ ($1-p$). 
        }
    	\label{fig:Baum}
\end{figure}

\noindent Equation (1) of the main text, which is used to fit the dependence of the excited state photoemission signal at \Kbar and \Kbarprime as a function of the circular polarization state of the pump pulse, was derived based on the binary tree-like model as illustrated in Fig.\ \ref{fig:Baum}. 
The total number of excited carriers at a given fluence of the pump laser light is assumed to be constant, but distributed between the \Kbar and \Kbarprime points, depending on the helicity of the light. The degree of circular polarization, $c = 0.5 \cdot \hat{S}_{3} + 0.5$ ($c^{\prime}=1-c$), is proportional to the intensity of the $\sigma^-$ ($\sigma^+$) component of the pump pulse and hence determines the fraction of the excited state population generated at \Kbar (\Kbarprime).
The spot size of the probe pulse is $\sim \SI{250}{\micro \meter}$ so that for typical domain sizes of the $\text{WS}_2$ layer in the low $\mu$m-range \cite{Bignardi2019} the trARPES signal results from averaging over a large number of individual domains. For a given fraction $f'=(1-f)$ of domains with mirror orientation (domain  orientation rotated by $\SI{60}{\degree}$ with respect to the preferential domain orientation $f$), a corresponding fraction of the probed excited state signal is due to photoemission from \Kbarprime. In a previous study on the structural properties of the used sample it was shown that $f'= \SI{0.00}{}$ with a maximum of \SI{0.05}{} contribution of mirror domains \cite{Bignardi2019}. The resulting limiting cases of $f = \SI{1}{}$ and $f = \SI{0.95}{}$ have been used as an input for the fit of Eq.~(1) of the main text to the experimental data.

For the quantitative analysis of the data we finally have to account for depolarization processes intrinsic to an individual $\text{WS}_2$ domain, such as deviations from the optical selection rules or intervalley scattering between \Kbar and \Kbarprime. These processes are considered by the parameter $p$, which gives the fraction of carriers behaving according to the optical selection rules. For the fits of Eq.~(1) to the experimental data, $p$ is kept as the only free fitting parameter.  
Following the branches in Fig.\ \ref{fig:Baum}, the transient ARPES intensity at \Kbar can be described by the following relation:
\begin{align}
     I_{\Kbar} \propto  \ &c \cdot f \cdot p \,+\, (1-c) \cdot (1-f) \cdot p \notag \\ 
     &+ \, c \cdot (1-f) \cdot (1-p)  \,+\, (1-c) \cdot f \cdot (1-p).
\end{align}
The relation describes the changes in the photoemission intensity due to photoexcitation at \Kbar. In order to describe the changes in the photoemission intensity at \Kbarprime, $c$ has to be replaced with $c' = 1-c$ due to the reversed optical selection rules.

\section{Rate equation model}

\noindent The characteristic time constants describing the decay dynamics of the transiently excited states probed at \Kbar were determined from a fit of a rate equation model to the time-resolved photoemission data. The temporal evolution of the photoemission intensities of conduction and valence band at \Kbar and \Kbarprime, $I_{\text{\Kbar},\text{CB}}$, $I_{\text{\Kbar},\text{VB}}$, $I_{\text{\Kbarprime}, \text{CB}}$, and $I_{\text{\Kbarprime},\text{VB}}$, are given by the following set of differential equations assuming a direct photoexcitation at \Kbar:
\begin{subequations}
    \begin{align}
        \frac{\partial I_{\text{\Kbar},\text{CB}}}{\partial t} &= - \frac{I_{\text{\Kbar},\text{CB}} - \alpha \cdot I_{\text{\Kbarprime} ,\text{CB}}}{\uptau_{\text{\Kbar}-\text{\Kbarprime}}} - \frac{I_{\text{\Kbar},\text{CB}}}{\uptau_{A}} + \left [ c \cdot p_{\text{VB}} + (1-c) \cdot (1-p_{\text{VB}})\right] \cdot g(t) ,
          \\
        \frac{\partial I_{\text{\Kbarprime},\text{CB}}}{\partial t} &= + \frac{I_{\text{\Kbar},\text{CB}} - \alpha \cdot I_{\text{\Kbarprime} ,\text{CB}}}{\uptau_{\text{\Kbar}-\text{\Kbarprime}}} - \frac{I_{\text{\Kbarprime},\text{CB}}}{\uptau_{A}} + 
        \left [1 - c \cdot p_{\text{VB}} - (1-c) \cdot (1-p_{\text{VB}})\right] \cdot g(t) ,
          \\
        \frac{\partial I_{\text{\Kbar},\text{VB}}}{\partial t} &= - \frac{I_{\text{\Kbar},\text{VB}}}{\uptau_{A}} + \left [ c \cdot p_{\text{VB}} + (1-c) \cdot (1-p_{\text{VB}})\right] \cdot g(t) ,
         \\
        \frac{\partial I_{\text{\Kbarprime},\text{VB}}}{\partial t} &= - \frac{I_{\text{\Kbarprime},\text{VB}}}{\uptau_{A}} + 
        \left [1 - c \cdot p_{\text{VB}} - (1-c) \cdot (1-p_{\text{VB}})\right] \cdot g(t).
    \end{align}
\end{subequations}

\noindent Here, $p_{\text{VB}}$ denotes the value for $p$ in the VB and is set to \SI{0.92}{}, as derived from the fits of Eq.~(1) to the results of the polarization scans for the UVB.
The optical excitation is described by a Gaussian pulse profile $g(t)$ with a FWHM of \SI{130}{\femto\s}. The excitation is distributed among \Kbar and \Kbarprime according to the circular polarization parameter $c = 0.5 \cdot \hat{S}_{3} + 0.5$. The decay of the excited state population in VB and CB is described by a characteristic time constant $\uptau_{A}$, which we associate with Auger-type scattering processes due to interaction with the gold substrate \cite{GrubisicCabo2015}. 
For the CB, the time constant $\uptau_{\text{\Kbar}-\text{\Kbarprime}}$ accounts additionally for intervalley scattering processes redistributing carriers between \Kbar and \Kbarprime. We exclusively consider spin-conserving intervalley scattering processes so that the energy of the relevant final CB state (momentum-forbidden dark state) for these types of processes is always lower than the energy of the directly photoexcited state [see Fig.\ 4(b) of the main text]. For a given valley selective excitation the factor $\alpha$ accounts for the resulting asymmetry in the intervalley scattering rate from \Kbar to \Kbarprime and vice versa. Fits to the experimental data showed that $\alpha$ stays consistently at a value of or close to zero. For the fitting results shown in Fig.~4(a) of the main text we therefore decided to keep $\alpha$ fixed to zero.

For the direct comparison of the rate equation model with the experimental data, the fraction of mirror domains contributing to the signal must in general be taken into account [$I(t) = f \cdot I_\text{\Kbar}(t) + (1-f) \cdot I_\text{\Kbarprime}(t)$]. According to the findings on the structural properties of the sample reported in Ref.~\cite{Bignardi2019} we set $f=\SI{1}{}$. The calculated intensities were finally scaled separately to match for the direct comparison with the experimental data.\\
\newpage
\noindent To avoid systematic errors potentially arising from a rotation of the sample, the analyzed data was obtained at the \Kbar point exclusively. The two data sets shown in Fig.\ 4(a) of the main text were recorded, however, with the polarization switched from predominantly $\sigma^-$ to $\sigma^+$ polarized light. For the fits with the $\sigma^+$ polarized pump, the expression $I_\text{\Kbarprime}(t)$ was used to simulate the scenario of a population via scattering from the directly excited valley.

\bibliography{bibliography}